\documentstyle[epsf,fleqn,aps]{revtex}

\def\Jrnl#1#2#3#4{{#1} {\bf #2}, #3 (#4)}

\def\PRB{Phys. Rev. B}

\def\PRL{Phys. Rev. Lett.}

\def\Section#1{}

\def\beq{\begin{equation}}

\def\eeq{\end{equation}}

\def\bea{\begin{eqnarray}}

\def\eea{\end{eqnarray}}

\def\age{\,\raise2pt\hbox{$\mathop{>}\limits_{\raise 2pt

\hbox{$\sim$}}$}\,}

\def\ale{\,\raise2pt\hbox{$\mathop{<}\limits_{\raise 2pt

\hbox{$\sim$}}$}\,}

\setbox122 = \hbox{1}

\def\id{\rlap{1}\rlap{\kern 1pt \vbox{\hrule width 4pt depth 0 pt}}

        \rlap{\kern 4 pt \hbox{\vrule height \ht122 depth 0 pt}}

           \hskip\wd122}

\begin{document}
\tolerance 50000
\twocolumn[\hsize\textwidth\columnwidth\hsize\csname
@twocolumnfalse\endcsname
\rightline{LPENS-Th-07/2001}
\title{Site-centered impurities in quantum spin chains}
\author{P.\ Pujol and J. Rech}

\address{
      Laboratoire de Physique,
      Groupe de Physique Th\'eorique \\
      ENS Lyon, 46 All\'ee d'Italie, 69364 Lyon C\'edex 07, France.\\}

\maketitle

\vspace{.5cm}

\begin{abstract}
The magnetic behavior of antiferromagnetic spin $1/2$ chains with
site-centered impurities in a magnetic field is investigated. The
effect of impurities is implemented by considering different
situations of both diagonal and off-diagonal disorder. The
resulting magnetization curves present a wide variety of plateaux,
whose position strongly depends on the kind of disorder
considered. The relevance of these results to experimental
situations is also discussed.

\vskip 0.5cm

PACS numbers: 75.10.Jm, \,75.10.Nr,\, 75.60.Ej.

\end{abstract}


\vskip -0.2cm
\vskip2pc]

\section{introduction}

The study of the magnetic properties of low dimensional
antiferromagnets has received much renewed attention during the
last years. One particular issue that captured both experimental
and theoretical efforts is the appearance of magnetization
plateaux in spin chains and ladders systems. In general, those
plateaux appear for pure spin systems at rational magnetization
values \cite{OYA,nos,Totsuka,Hida}. Some experiments have indeed
confirmed these theoretical predictions in a few particular cases
\cite{S1dim}.

More recently, the effect of the presence of impurities on such
magnetic behavior was also investigated. From the theoretical
point of view, the properties at zero magnetic field have largely
been elucidated \cite{FXXZ,HG}, and recent analysis explored also
the robustness of the rational plateaux for small amplitude of the
disorder \cite{tot}. A very interesting phenomenon recently
discovered was also that, for bond-disorder with discrete
probability distributions, plateaux at non-rational values of the
magnetization are present \cite{cdgpp}. Moreover, the position of
such plateaux in the curve is related in a simple way to the
concentration of impure bonds in the system. Since this kind of
disorder is a good candidate for modelling concrete experimental
situations, this result opens new perspectives in the
interpretation of experimental curves. In order to make closer
contact with experiments, one has to take into account all the
effects that can be produced by non-magnetic impurities, and
provide a wide range of possibilities for implementing the
presence of such impurities in realistic models.

In this paper we analyze a simple model where different kinds of
diagonal and off-diagonal disorder are present. The techniques
used are based on a decimation procedure \cite{DM}, as well as
standard quantum mechanics tools and numerical calculations for
$XX$ chains of sizes ranging from 50 000 to 100 000 sites. Despite
the simplicity of our model, we find that a huge number of
families of cases is present, each of them presenting its own
hierarchy of magnetization plateaux. Moreover, we show how a
simple tuning of the microscopic parameters can drive our system
from one family to another. We also argue that this simple example
serves as a good laboratory for understanding and classifying the
wide variety of cases one can encounter in experimental curves of
antiferromagnetic systems with impurities.

\section{The model}

 The model we are going to study is a spin-1/2
chain whose Hamiltonian is given by:
\beq
\hspace{-0.7cm}
\label{Hamiltonian}
H = \sum_i \left[ J_i\, \left( S^x_i S^x_{i+1} + S^y_i S^y_{i+1} +
\Delta S^z_i S^z_{i+1}\,\right)\, + h_i\, S^z_i \right] .
\eeq

The randomness is implemented by considering $J_i$ and $h_i$ as
random variables. Specifically, bond randomness will be obtained
by assigning to all the variables $h_i$ the same value fixed at
$h$, and by giving a probability distribution to the set of
variables $ \{J_i \} $. The probability distribution can, of
course, be chosen in such a way to give, on average, a periodicity
of $q$ to the bond variables. It is important to say that
realistic distributions relevant for many possible experimental
situations are discrete. In this sense, a binary distribution
captures the essential characteristics of the phenomena we want to
describe here, the generalization to more complicated discrete
distributions being straightforward. In \cite{cdgpp}, a
distribution of the form:
\beq
\hspace{-0.7cm}
\label{q-bin}
P(J_i) =  p \: \delta(J_i-J') + (1-p)\: \delta(J_i-J - \gamma_i
J)\,,
\eeq
was considered, where $\gamma_i \equiv \gamma,\: (-\gamma)\,$ if
$i = q n,\:$ ($ i \ne q n\,$). In order to match the behavior of a
disorder originating from a site impurity, a most appropriate
distribution can be given by the following algorithm: distribute
first regular values for the bonds $J_i$ with the desired
periodicity $q$, and parameter $\gamma$, then, to each site of the
chain assign a probability $1-p$ to be an "original" site, and $p$
to be an "impure" site. Finally, for each site $i$ which turned
out to be impure, change the values of the surrounding bonds
$J_{i}$ and $J_{i-1}$ to $J'$. We are, as in \cite{cdgpp},
concerned with three different values for the bond strength, $J
\pm \gamma J$ and $J'$, but now with a correlated probability
distribution.

For site-centered disorder, we are going to consider two different
cases. The first is just obtained by writing $h_i$ as  $h + h'_i$
where $h'_i$ is a random variable taking values of $0$ with
probability $p$ and $h'$ with probability $1-p$. This case can be
considered as academic, since it violates the symmetry $ h
\rightarrow -h$, but it will nevertheless provide useful insight
for more realistic disorders. The second and more realistic case
is the $ h \rightarrow -h$ symmetry preserving case where $h_i = h
(1 + \alpha_i)$, with values for $\alpha_i$ being binary
distributed among $0$ and $\alpha$ with probability $p$ and $1-p$,
and $\alpha \geq -1$ (the limiting case of $\alpha = -1$
corresponding to an impurity which does not couple to the magnetic
field).

In the case of $\Delta =0$, the model can be mapped, via the well
known Jordan-Wigner transformation, to a problem of free fermions,
whose first quantized Schr\"odinger-like equation is:
\beq
J_{i-1}~c_{i-1} + J_{i}~c_{i+1} = h_i~c_i .
\eeq

The magnetization is simply related to the number of states
occupied and the $z$ component of the susceptibility is just
proportional to the density of states. For a given energy, the
number of states can be simply obtained by the node counting
method \cite{er}, related to the number of positive "self-energy"
variables:
\beq
\Delta_i = c_{i-1}~J_{i-1}/ c_i .
\eeq
The recursive formula for these variables is:
\beq
\Delta_{i+1} = J_i^2/(h_i -\Delta_i)
\eeq
which, for the case of bond and $ h \rightarrow -h$ symmetry
preserving site disorder can be written as:
\beq
D_{i+1} = W_i^2/(h -D_i)
\label{changevar}
\eeq
where $D_i = \frac{\Delta_i}{1+\alpha_i}$ and $W_i^2 =
\frac{J_i^2} {(1+\alpha_i)(1+\alpha_{i+1})}$. Note that $D_i$ has
the same sign as $\Delta_i$, then the node counting can be done in
the $D$ variables. This remark will be very useful when discussing
the presence of plateaux and the behavior of the magnetization
curve close to $h=0$.

\section{Off-diagonal disorder}

\begin{figure}
\hbox{%
\epsfxsize=3.2in \hspace{-0.5cm} \epsffile{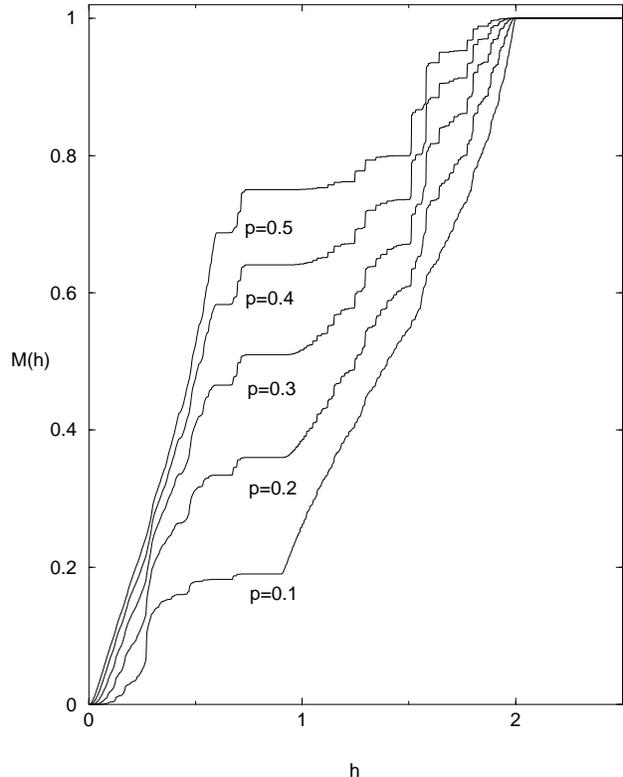}}
\caption{Magnetization curves of $XX$ chains with off-diagonal
disorder for the parameters values of $J=2$, $J'=0.6$,  $\gamma
=0.45$.
\label{fig1} }
\end{figure}

As mentioned above, this case corresponds to a generalization for
bond-disorder considered in \cite{cdgpp}. In this case each time a
site is considered as an impurity, it must be surrounded by bonds
with lower values. The decimation procedure used in \cite{cdgpp}
is again well adapted for identifying the location of
magnetization plateaux. However, some care has to be taken in
counting the number of bonds for each decimation step,
particularly in the case of two impurities sitting in neighboring
sites. We refer to the reference mentioned above for the details
of this reasoning and present here the result for a dimerised
chain which is the most relevant case for experimental situations,
the generalization to generic periodicity of the lattice being
straightforward. The principal plateau is located at:
\beq
M = 2p - p^2
\eeq
coming from the decimation of the bonds with the highest value.
The second step of the decimation gives a secondary plateau at:
\beq
M  = 2p - 2p^2 + 2p^3 - p^4.
\eeq
These plateaux are well observed in the numerical curves of figure
(\ref{fig1}), where we show the result for an $XX$ chain.

The behavior of the magnetization for small magnetic fields can
also be obtained as a generalization of the normal off-diagonal
disorder. For example, for the $XX$ case, a mapping to a random
walk in the self-energy variables shows, as for a standard
dimerised chain with random bonds \cite{rusos,HG}, a power law
behavior for the magnetic susceptibility for even periodicity of
the lattice:
\beq
\label{chi-even}
\chi_z \propto  H^{\lambda -1}\,
\eeq
where the exponent can be easily obtained from the mean and
variance of $\ln (J)$ (see for example \cite{cdgpp}). Using the
same method, a logarithmic behavior for the susceptibility is
obtained for odd periodicity of the lattice:
\beq
\label{chi-odd}
\chi_z \propto \frac{1}{H[\ln(H^2)]^3}.
\eeq

\section{Diagonal disorder}

\begin{figure}
\hbox{%
\epsfxsize=3.2in \hspace{-0.5cm} \epsffile{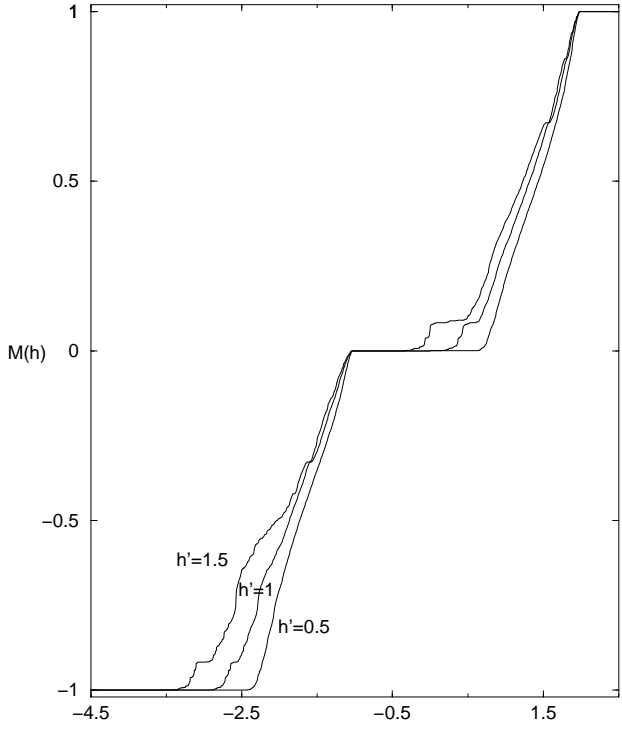}} \caption{
Magnetization curves for the first kind of diagonal disorder
mentioned in the text (see eq. (\ref{dia1}) taken with $p=0.3$,
$\gamma =0.5$, $J=2$.
\label{fig2} }
\end{figure}

\begin{figure}
\hbox{%
\epsfxsize=3.2in \hspace{-0.5cm} \epsffile{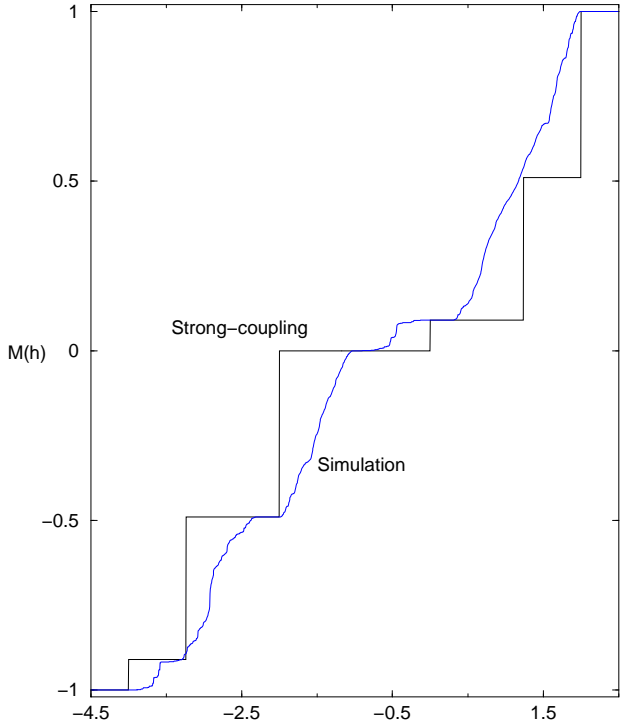}}
\caption{Same as in figure (\ref{fig2}) with $J=2$, $\gamma=0.45$,
$h'=2$, $p=0.3$.
\label{fig3} }
\end{figure}

We first concentrate on the naive diagonal disorder one can
introduce in a dimerised chain, by supposing that a supplementary
magnetic field $h'$ at each impurity is present. While for low
$h'$ no noticeable changes at the curve occur, for strong enough
values of $h'$, we see in figure (\ref{fig2}) the appearance of
new plateaux. A simple way to understand such plateaux is by
considering the strong coupling case $\gamma \sim 1$. The order
zero in powers of $(1-\gamma)$ is just given by a combination of
dimers which can contain two, one or zero impurities with
probability $(1 -p)^2/2$, $p(1-p)$ and $p^2/2$ respectively. It is
then easy to draw the magnetization curve for each case (a three
step stair like curve) and then superpose the curves with the
appropriate weight. The result is compared to the numerical data
in figure (\ref{fig3}). The positions of the plateaux predicted at
this order are located at:
\beq
M = - 1+p^2 , ~ - (1-p)^2 , ~ 0 ,
 ~ p^2 , ~ 2p-p^2.
\label{dia1}
\eeq

Note in passing the lack of $M \leftrightarrow -M$ symmetry in the
magnetization curve due to the very nature of the disorder.
Comparing with the numerical result, one sees that some plateaux
are softened due to the presence of non-zero coupling which was
neglected in our procedure. The next step is as usual to turn on a
non-zero value for $\gamma$ and use standard quantum mechanics
perturbation theory (see for example \cite{cdhps} for a similar
treatment). The result is a smoothing of the jumps between the
plateaux and a correction to their width which can be calculated
in powers of $(1-\gamma)$. Since we are concerned with more
realistic kinds of disorder these computations are beyond the
scope of this work. We just mention the fact that this case leaves
the $M=0$ plateau present, a result which can be also obtained for
the $XX$ case by the random walk argument. We refer to \cite{er}
for the details of this mapping, and just mention here that it is
easy to study the cycles of $+-$ self energy variables for low
energies. One can indeed show that, in the equivalent first
quantized Schr\"odinger picture, and for small enough but non-zero
values of the energy, these cycles are infinite indicating that
the number of states remains constant in a finite interval of
energy, which translate in the magnetic language into a zero
magnetization plateau.

\begin{figure}
\hbox{%
\epsfxsize=3.2in \hspace{-0.5cm} \epsffile{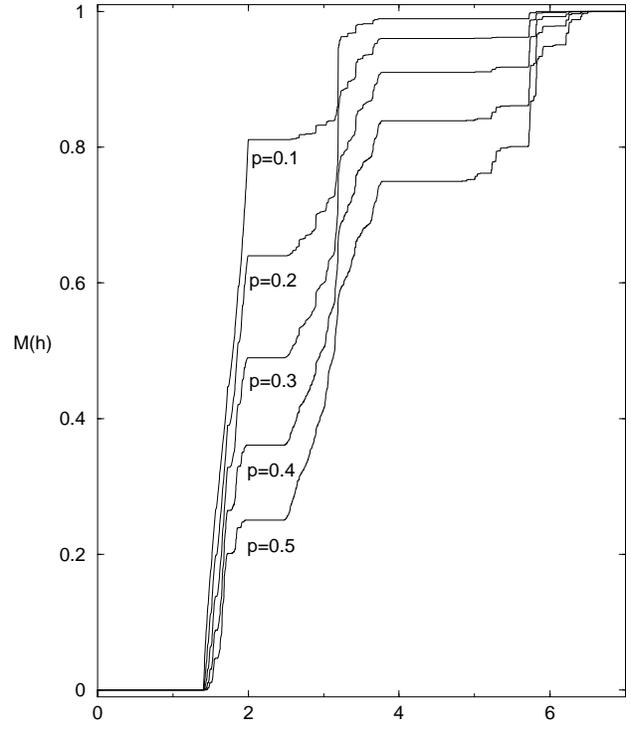}}
\caption{Magnetization curves for the $(h \leftrightarrow -h)$
symmetric disorder with $J=2$, $\alpha =-0.7$, $\gamma=0.7$.
\label{fig4} }
\end{figure}

The second case is obtained by assuming that each impurity couples
to the magnetic field with a factor of $(1 +\alpha)$. It gives a
symmetric magnetization curve by the change $h \rightarrow -h$. It
is sufficient to study only the case $\alpha < 0$. Indeed, the
case of positive $\alpha$ is obtained by a simple re-scaling of
the magnetic field and the change $ p \leftrightarrow 1-p$. In the
case of $XX$ chains, we can make use of the change of variables
defined in (\ref{changevar}). We have then the equivalence between
the model we are considering and a system with constant magnetic
field, but with a wide hierarchy of values for the bonds. The
decimation procedure can then be used to predict the presence of
plateaux. We have however to identify different cases
corresponding to each hierarchic structure of the bonds, and these
are given by the values of $\alpha$ and $\gamma$ we are
considering. As already stated before, the decimation procedure is
implemented by eliminating the strongest bonds, and their relative
weight in the system gives the position of the corresponding
plateau in magnetization. We omit the details of the combinatorics
and present the results for the position of the principal plateaux
for each case in table (\ref{tab1}). Figure (\ref{fig4}) shows
also the magnetization curve obtained for one of these cases,
showing a fair agreement with the decimation predictions.

Let us now analyze the behavior of the system at low magnetic
fields. By using again the mapping to a random walk of the
generalized Riccati variables, one could naively say that equation
(\ref{changevar}) gives rise to a behavior similar to the one
obtained for an off-diagonal disorder. An important difference is
however that the noise in this case is cancelled step by step,
giving rise to a regular and non-random walk at large times (this
can be easily seen by inspecting the form of the variables $W_i$).
Then, at low magnetic fields, the behavior of our systems is
asymptotically equivalent to the one of a pure chain, and for an
even periodicity of the bonds shares in particular the presence of
a plateau at zero magnetization. Strictly speaking, these results
are valid for $XX$ chains only, since we have used the change of
variables (\ref{changevar}). We conjecture however that the
qualitative behavior for the curve, and in particular, the
position of the magnetization plateaux remains valid for generic
$XXZ$ chains.

\section{Mixed disorder}

This last case can be considered as the combination of both kinds
of disorder studied before. It is important to notice that the
equivalence between positive and negative $\alpha$ is not valid
anymore due to the extra presence of bond disorder. For the $XX$
case, it is however easy to see by means of (\ref{changevar}) and
the subsequent decimation procedure, that the case $\alpha >0$ has
the same plateaux as for the off-diagonal disorder discussed
above. This is still the case for negative but small values of
$\alpha$, since the hierarchy of coupling in the effective system
is unchanged. The situation is however radically changed when
$\alpha$ goes below the critical value:
\beq
\alpha_c = { J' \over J + \gamma J} -1
\eeq
where the hierarchy of values for the bonds is completely changed.
Figure (5) shows that crossing the critical value of $\alpha$
indeed changes completely the nature of the magnetization curve
and the position of the plateaux in the curve. In that case, the
strongest bonds have the values $J'/(1 + \alpha)$ and can be found
in arbitrary long chains of impurities. This fact invalidates the
standard decimation procedure for locating the position of the
plateaux. However, one can proceed by noticing that, for values of
$\alpha$ close enough  to $-1$, the inter-impurities bonds in the
effective model are much larger than the others. One can then use
a kind of strong-coupling treatment in which the zeroth order
system is obtained by equating the value of the remaining bonds to
zero. We are then left with a collection of decoupled spin chains
with arbitrary length. For an array containing $n$ spins, the
eigenvalues of the Hamiltonian are simply given  by the roots of
second order Chebyshev polynomials
\beq
E_p = J_{eff} ~cos( {p \pi \over n+1})~~;~~ p=1...n
\eeq
and $J_{eff}$ is the value of the coupling between the spins. The
magnetization curve for such a finite chain can also be easily
obtained. To draw the magnetization curve for the total system one
just has to superpose the curves for all possible values of $n$
with the corresponding probability of appearance $(1-p)^2 p^n$.
The result is shown in figure (\ref{fig6}) where we compare the
numerical result with the strong coupling computation performed up
to 11th order in $p$. The appearance of a hierarchy of plateaux
close to saturation predicted by the theory is clearly observable
in the numerical data.

The plateau at low values of the magnetization and the jump in the
magnetization at zero field is an artifact of the zeroth order
approximation. This can be cured now by turning on the remaining
coupling to non-zero values and treating them in perturbation
theory. On what concerns the jump at $M=0$, there is however a
much simpler way of studying the shape of the curve close to
$M=0$. This is achieved again by the mapping to the random walk
problem. One recovers in this way the power law  or logarithmic
behavior as in (\ref{chi-even})-(\ref{chi-odd}) depending on the
periodicity of the bonds.

\section{conclusion}

To summarize, we have shown that the magnetization curve of spin
chains with different kinds of impurities exhibits a rich variety
of plateaux. While the phenomenon of non-rational plateaux was
already observed in a disordered case \cite{cdgpp}, we have shown
in this paper with simple examples that the position of such
plateaux strongly depends on the particular values of the
microscopic parameters. While one can speak of a sort of
universality for the position of the plateaux, depending only on
the concentration of impurities and periodicity of the averaged
bonds, we have clearly seen that different families of disorder
produce qualitatively different magnetization curves. Moreover,
within the same kind of disorder (the mixed disorder case in
particular) a simple tuning of the microscopic parameters can
switch the system from one "universality class" to another. This
scenario teaches us that the characterization of spin chains with
impurities is better understood in terms of families of disorder,
at least to understand the behavior of the magnetization curve.
While some results in this paper have been obtained for the $XX$
case only, it is reasonable to think that such a characterization
in terms of families of disordered chains remains valid for
generic $XXZ$ chains. Of particular interest is the mixed case,
for which we believe the case $\alpha \rightarrow -1$ should
represent the magnetization curve for materials such as CuGeO$_3$
doped with Si \cite{SiCuGeO}. We trust this work provides the
necessary tools for predicting the shape of the magnetization
curve in future experimental situations, where higher
periodicities than $2$ are also conceivable \cite{Hida}.

We acknowledge B. Dujardin for fruitful discussions and D. C.
Cabra and A. Honecker for constructive comments and a careful
reading of the manuscript.



\vspace{0.6cm}

\begin{figure}
\hbox{%
\epsfxsize=3.2in \hspace{-0.5cm} \epsffile{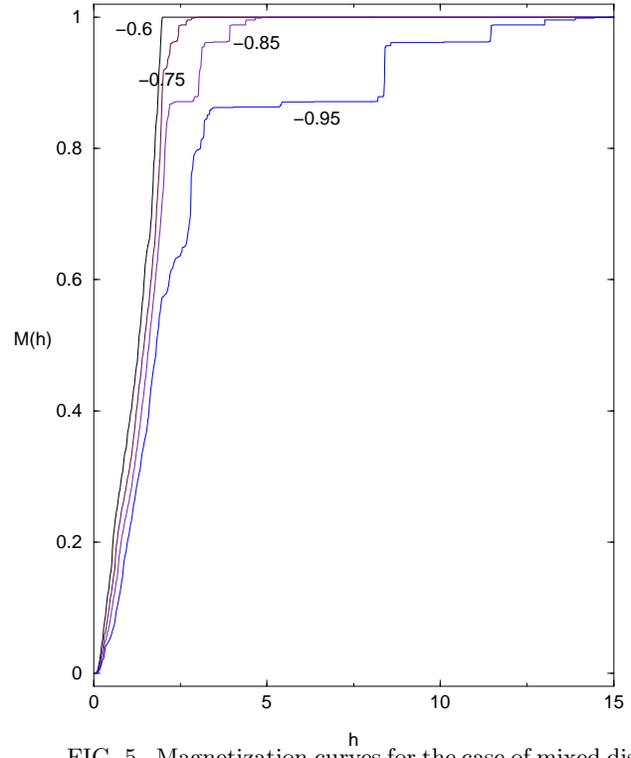}}
\caption{Magnetization curves for the case of mixed disorder with
$J=2$, $J'=0.8$, $\gamma=0.5$, $p=0.4$ and different values of
$\alpha$. In this case $\alpha_c =-0.73 $ (see text).
\label{fig5} }
\end{figure}


\begin{figure}
\hbox{%
\epsfxsize=3.2in \hspace{-0.5cm} \epsffile{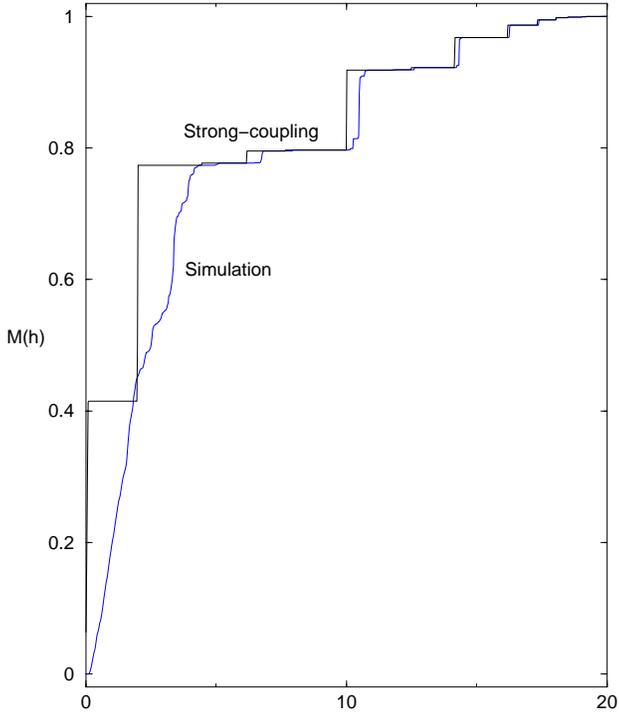}}
\caption{Strong-coupling and numerical result for the mixed
disorder case with $J=2$, $J'=0.8$, $\alpha=-0.95$, $\gamma=0.5$,
$p=0.3$.
\label{fig6} }
\end{figure}


\begin{table}
\begin{tabular}{|c|c|}
\hline
Range of $\alpha$ parameter & Magnetization plateaux \\
\hline
&  $1-p^2$ \\
$\big(\frac{1-\gamma}{1+\gamma}\big)^2-1 < \alpha < 0$ & $(1-p)^2$ \\
 & $0$ \\
\hline
 &  $1-p^2$ \\
 & $1-p^2-p^2(1-p)^2$ \\
$\big(\frac{1-\gamma}{1+\gamma}\big)^2-1 < \alpha < \big(\frac{1-\gamma}{1+\gamma}\big)^4-1$ & $(1-p)^2(1+p^2)$ \\
 & $(1-p)^2$ \\
 & $0$ \\
\hline
 & $1-p^2$ \\
 & $1-p^2-p^2(1-p)^2$ \\
 & $1-p^2-p^2(1-p)^2(1+p^2)$ \\
$\alpha < \big(\frac{1-\gamma}{1+\gamma}\big)^4-1$ & $(1+p^2+p^4)(1-p)^2$ \\
 & $(1+p^2)(1-p)^2$ \\
 & $(1-p)^2$ \\
 & $0$ \\
\hline
\end{tabular}
\caption{ Magnetization at the principal plateaux for the $(h
\leftrightarrow -h)$ symmetric diagonal disorder (see text).
\label{tab1} }
\end{table}

\end{document}